\documentclass[journal, ,x11names]{IEEEtran}
\usepackage{setspace}

\usepackage{subfigure}
\usepackage{amsmath,amsfonts}
\usepackage{algorithmic}
\usepackage{algorithm}
\usepackage{array}
\usepackage{subfig}
\usepackage{textcomp}
\usepackage{stfloats}
\usepackage{url}
\usepackage{verbatim}
\usepackage{graphicx}
\usepackage{cite}
\usepackage{subfloat}
\usepackage{tikz}
\usepackage{bm}

\usepackage{color,xcolor}
\usepackage{caption}
\usepackage{pgfplots}
\usetikzlibrary{calc,arrows.meta,positioning}
\usetikzlibrary{shapes.arrows}
\usetikzlibrary{fit,backgrounds} 
\usepackage{standalone}
\usepackage[numbers,sort&compress]{natbib}
\usepackage{setspace}
\usepackage{makecell}

\begin{document}
\title{Contextual Wireless Video Semantic Communication in MIMO-OFDM Systems}

\author{Bingyan Xie, Cong Zhou, Yuxuan Shi, Biqian Feng, Yongpeng Wu, Wenjun Zhang,~\IEEEmembership{Fellow,~IEEE}
        % <-this % stops a space
\vspace{-15pt}

\thanks{The work of Yongpeng Wu was supported in part by the National Key Research and Development Program of China under Grant 2022YFB2902100; in part by the Fundamental Research Funds for the Central Universities; in part by the National Science Foundation under Grant 62122052 and Grant 62071289; in part by the 111 Project under Grant BP0719010; and in part by STCSM under Grant 22DZ2229005. (Corresponding author: Yongpeng Wu.)}
\thanks{Bingyan Xie, Yongpeng Wu, Biqian Feng, and Wenjun Zhang are with the Department of Electronic Engineering, Shanghai Jiao Tong University, Shanghai 200240, China (e-mail: bingyanxie, yongpeng.wu, fengbiqian, zhangwenjun@sjtu.edu.cn).}
\thanks{Cong Zhou is with the School of Electronics and Information Engineering, Harbin Institute of Technology, Harbin, 150001, China. (e-mail: zhoucong@stu.hit.edu.cn).}
\thanks{Y. Shi is now with the Department of Networked Intelligence, Peng Cheng Laboratory, Shenzhen 410083, China (e-mail: shiyx01@pcl.ac.cn).}

}
%{Shell \MakeLowercase{\textit{et al.}}: A Sample Article Using IEEEtran.cls for %IEEE Journals}

%\IEEEpubid{IEEE}

\maketitle

\begin{abstract}
This paper proposes a MIMO-OFDM-based context video semantic transmission framework, namely M-CVST, for robust video communication over multi-path multiple-input multiple-output (MIMO) channels. It introduces a context-subcarrier correlation map that aligns video feature context with groups of MIMO subcarriers. To leverage the time-correlated nature of multi-path channels, a recursive subcarrier sampling method paired with time-correlated reference embedding is designed, enabling the use of previously sampled MIMO subcarrier CSI to enhance channel state awareness in the entropy coding model. Numerical results verify the superiority of proposed M-CVST over MIMO multi-path channels compared to other semantic schemes and traditional separated schemes.

\end{abstract}

\vspace{-10pt}
\section{Introduction}\label{s1}
\IEEEPARstart{T}he proliferation of video-centric applications (e.g., virtual reality, Internet of Vehicles, smart cities) dominates Internet traffic, imposing heavy pressure on wireless transmission systems. Separated source-channel coding (SSCC), which combines H.265/VVC \cite{265, vvc} with low density parity check (LDPC), is widely adapted. However, deep learning-based joint source-channel coding (JSCC) delivers superior performance under finite blocklength conditions \cite{wvscd,cra,dvsc,dvst,cvst}, inspiring various video-oriented semantic communication frameworks. Xie et al. \cite{wvscd} proposed a semantic-level approach for efficient frame correlation modeling; Niu et al. designed signal-to-noise ratio (SNR)-adaptive channel coding with semantic restoration; Wang et al. \cite{dvst} developed a context-based nonlinear transform coding (NTC) framework for variable-length transmission. Overall, context-aware transmission schemes \cite{dvst,cvst} outperform residual-based methods \cite{dvsc} in compression and support multi-reference awareness.

Although \cite{cvst} considered multi-input multi-output (MIMO) channel scenarios, it primarily assumes simple i.i.d. conditions, overlooking more sophisticated practical multi-path fading and time correlation characteristics. Orthogonal frequency division multiplexing (OFDM) is widely adopted to mitigate multi-path effects: it splits high-speed data streams into multiple parallel subcarriers, extending symbol duration beyond the typical delay spread and converting frequency-selective fading channels into flat-fading subchannels. In time-varying scenarios, wireless channels present time-correlated variations, posing a critical challenge: how to effectively utilize historical channel information to enable transmission frameworks to better estimate and adapt to current channel states.

Based on the above insights, we propose M-CVST, a MIMO-OFDM integrated context-aware video semantic transmission framework for multi-path MIMO channels. Inspired by the context-channel correlation map \cite{cvst}, M-CVST establishes fine-grained correlations between feature context and wireless channels at the subcarrier level. To reduce the overhead of channel state information (CSI) acquisition and precoding, we design a recursive subcarrier sampling method that periodically samples CSI from one subcarrier in each subcarrier group across successive OFDM symbols. We also aggregate previously learned correlation maps as temporal references to boost entropy coding performance. Our main contributions are summarized as follows:

\begin{enumerate}
\item{\textbf{M-CVST Framework:}}
A context-aware video semantic transmission framework for robust transmission over practical multi-path MIMO-OFDM channels is proposed. Unlike \cite{cvst}  which aligns feature contexts with entire MIMO subchannels, it constructs a context-subcarrier correlation map at the individual subcarrier granularity, enabling customized designs to tackle multi-path propagation challenges in MIMO systems.
\item{\textbf{Recursive Subcarrier Sampling:}}
A recursive subcarrier sampling method is designed to cut CSI acquisition and precoding computation overhead. Successive positions are sampled across consecutive OFDM symbols in each group, allowing collection of prior CSI from other subcarriers for subsequent processing.
\item{\textbf{Time-Correlated Reference Embedding:}}
A time-correlated reference embedding is proposed to integrate channel time variability into entropy coding. Converting pre-learned channel-subcarrier correlation maps into this embedding empowers the entropy model to better adapt to the instantaneous states of multi-path channels.
\end{enumerate}

\begin{figure*}[htbp]
	\centering
	\includegraphics[width=6.1in]{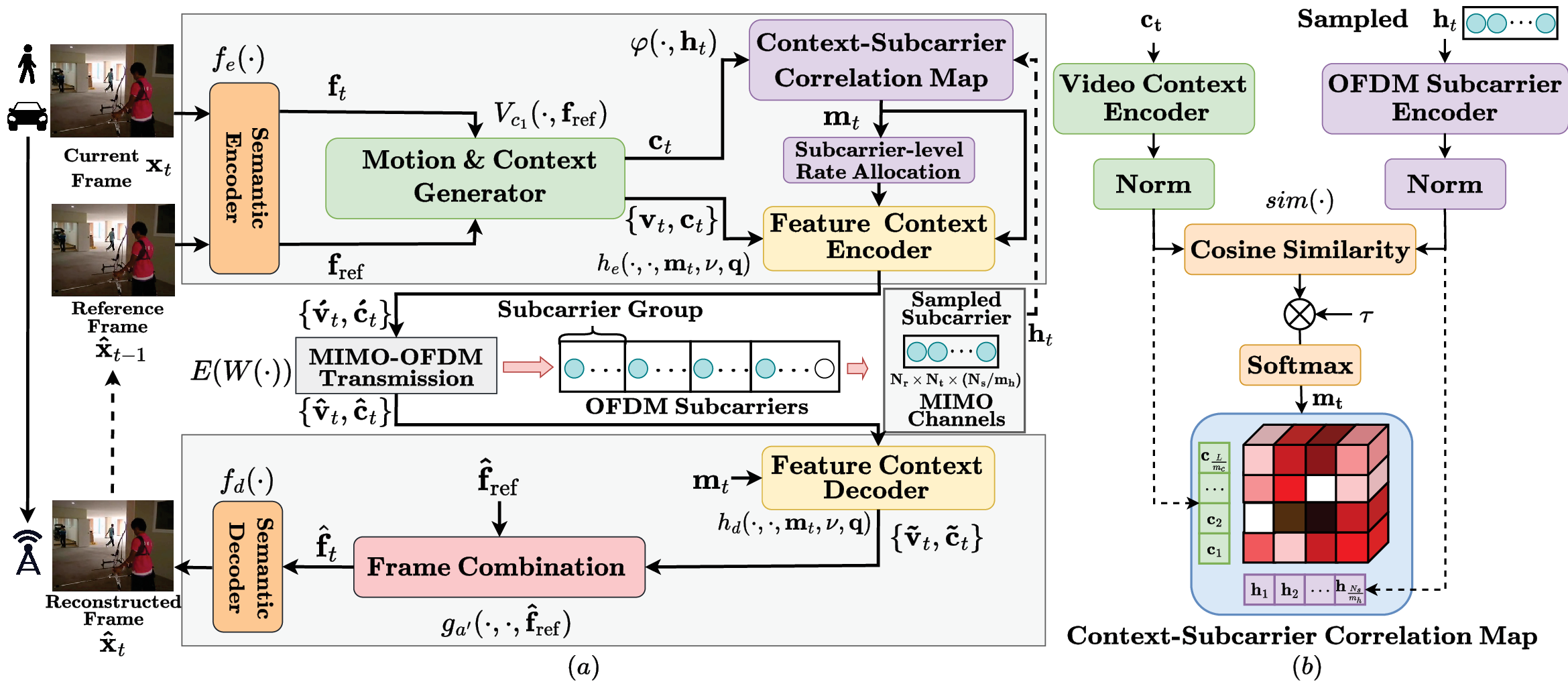}
	\vspace{-5pt}
	\caption{(a) The proposed M-CVST framework for uplink wireless video transmission from vehicles/pedestrians to base stations. (b) The structure of context-subcarrier correlation map.}
	\vspace{-15pt}
	\label{fig_1}
\end{figure*}

Notational Conventions: $\mathbb{R}$ and $\mathbb{C}$ refer to the real and complex number sets, respectively. $\mathcal{N}\left (\mu, \sigma^2 \right)$ denotes a Gaussian distribution with mean $\mu$ and variance $\sigma^2$. $\lfloor\cdot \rceil$ denotes the quantization operation. $\left(\cdot\right)^{H}$ denotes the Hermitian, $\left(\cdot\right)^{-1}$ is the matrix inverse, $\log(\cdot)$ denotes the logarithm operation.

\vspace{-10pt}
\section{Proposed M-CVST Framework}

The proposed M-CVST framework is shown in Fig. \ref{fig_1}(a). For arbitrary GoP $\mathbf{X} = \left\{\mathbf{x}_1,\mathbf{x}_2,\cdots,\mathbf{x}_T\right\}$, it contains $T$ successive frames with $\mathbf{x}_t\in \mathbb{R}^{3\times H\times W}$, $t=1,2,\cdots, T$. For the inter-coded frame (P frame) with $t=2,\cdots,T$, the semantic encoder, $f_e(\cdot): \mathbb{R}^{3\times H\times W}\longrightarrow \mathbb{R}^{L\times H'\times W'}$, encodes $\mathbf{x}_t$ into the semantic features $\mathbf{f}_t$. Then, the motion vectors $\mathbf{v}_t\in \mathbb{R}^{L\times H'\times W'}$ and the context $\mathbf{c}_t\in \mathbb{R}^{L\times H'\times W'}$ are learned through the motion $\&$ context generator $V_{c1}(\cdot,\mathbf{f}_{\mathrm{ref}}):\mathbb{R}^{L\times H'\times W'}\longrightarrow \mathbb{R}^{L\times H'\times W'}\times\mathbb{R}^{L\times H'\times W'}$, where $\mathbf{f}_{\mathrm{ref}}=\mathbf{f}_{t-1}$.

With both learned feature context and feedback sampled MIMO-OFDM subcarriers, we generate a context-subcarrier correlation map $\mathbf{m}_t \in \mathbb{R}^{(L/m_c)\times (N_s/m_h)}$ through $\varphi(\cdot,\mathbf{h}_t):\mathbb{R}^{L\times H'\times W'}\longrightarrow\mathbb{R}^{(L/m_c)\times (N_s/m_h)}$ to represent their correlation, where $m_c$ is channel dimension of the feature context group, $m_h$ is subcarrier number in a subcarrier sampling group, $\mathbf{h}_t\in\mathbb{R}^{N_r\times N_t\times (N_s/m_h)}$ is the feedback $N_r\times N_t$ MIMO CSI sampled in $N_s$ subcarriers. Then $\mathbf{v}_t$ and $\mathbf{c}_t$ are fed into the feature context encoder $h_{e}(\cdot, \cdot, \mathbf{m}_t,\nu,\mathbf{q}): \mathbb{R}^{L\times H'\times W'}\times\mathbb{R}^{L\times H'\times W'}\longrightarrow \mathbb{R}^{L_{v}}\times \mathbb{R}^{L_{c}}$ to achieve the CSI-aware semantic coding for robust and flexible semantic codewords generation, $\mathbf{\acute{v}}_t \in \mathbb{R}^{L_{v}}$ and $\mathbf{\acute{c}}_t \in \mathbb{R}^{L_{c}}$, in terms of various channel bandwidth ratios (CBRs). $L_{v}$ and $L_{c}$ are the respective final transmitted video codeword lengths. $\nu$ refers to the SNR value, while $\mathbf{q}$ refers to rate adaptive terms.

To tackle the effect brought by multi-path MIMO channels, time-frequency domain transition is adopted. This allows us to leverage OFDM to mitigate frequency-selective fading, where each OFDM symbol encompasses a MIMO channel denoted as $\mathbf{H}_t=[\mathbf{H}_{t,1},\cdots,\mathbf{H}_{t,N_s}] \in \mathbb{C}^{N_r \times N_t \times N_s}$. To alleviate the substantial overhead associated with full CSI acquisition and singular value decomposition (SVD) precoding, we partition the subcarriers into $m_h$ groups, each containing $N_s/m_h$ adjacent subcarriers. Within every group, only one representative subcarrier is sampled, and its CSI is used to approximate the channel response for the entire group.

$\mathbf{\acute{v}}_t$ and $\mathbf{\acute{c}}_t$ are then reshaped and precoded by the SVD as
\begin{align}
	\{\mathbf{\hat{v}}_{t,i},\mathbf{\hat{c}}_{t,i}\}=\Lambda_{t,i}^{-1}\mathbf{U}_{t,i}^H\mathbf{H}_{t,i}\mathbf{V}_{t,i}\{\mathbf{\acute{v}}_{t,i},\mathbf{\acute{c}}_{t,i}\}+\Lambda_{t,i}^{-1}\mathbf{U}_{t,i}^H\mathbf{n}, 
\end{align}
where $\mathbf{n}$ is the complex Gaussian channel noise vector whose component has zero mean and covariance $\sigma^{2}$. SVD decomposes the sampled MIMO channel matrix $\mathbf{h}_{t,i}$. $\mathbf{h}_{t,i}=\mathbf{U}_{t,i}\Lambda_{t,i}\mathbf{V}_{t,i}^H$ with $\mathbf{U}_{t,i}\in \mathbb{C}^{N_r\times N_r}$, $\mathbf{V}_{t,i}\in \mathbb{C}^{N_t\times N_t}$ and $\Lambda_{t,i}\in \mathbb{R}^{N_r\times N_t}$ for the $i$-th subcarrier.

At the receiver, with the feature context decoder $h_{d}(\cdot, \cdot,\mathbf{m}_t,\nu,\mathbf{q}): \mathbb{R}^{L_{v}}\times\mathbb{R}^{L_{c}}\longrightarrow \mathbb{R}^{L\times H'\times W'}\times\mathbb{R}^{L\times H'\times W'}$, received motion vector and context are translated to $\mathbf{\tilde{v}}_t$ and $\mathbf{\tilde{c}}_t$ with the help of $\mathbf{m}_t$. Since $\mathbf{m}_t$ only costs minor transmission rate, it is losslessly shared to the decoder. Through the frame combinator $g_a'(\cdot,\cdot,\mathbf{\hat{f}}_{\mathrm{ref}}):\mathbb{R}^{L\times H'\times W'}\times\mathbb{R}^{L\times H'\times W'}\longrightarrow \mathbb{R}^{L\times H'\times W'}$, semantic frame $\hat{\mathbf{f}_t}$ is reconstructed. Finally, the semantic decoder, $f_d(\cdot):\mathbb{R}^{L\times H'\times W'}\longrightarrow \mathbb{R}^{3\times H\times W}$, converts $\hat{\mathbf{f}}_t$ into $\hat{\mathbf{x}}_t$ and outputs the final reconstructed GoP $\hat{\mathbf{X}}=\left\{\hat{\mathbf{x}}_1,\hat{\mathbf{x}}_2,\cdots,\hat{\mathbf{x}}_T\right\}$ frame by frame.
\vspace{-10pt}
\section{Context Channel Correlation Map in Subcarrier Level}
\cite{cvst} proposed a context-channel correlation map that aligns hierarchical video semantic context with single-carrier MIMO subchannel quality for unequal error protection (UEP). Yet this subchannel-level mapping is inapplicable to practical multi-path MIMO-OFDM systems where subcarriers exhibit heterogeneous transmission characteristics due to multi-path fading. To address this, we design a context-subcarrier correlation map that achieves quality-aware hierarchical alignment between semantically critical video feature context and each MIMO-OFDM subcarrier’s transmission properties. Since subcarrier channel capacity directly quantifies transmission reliability and video semantic context has differentiated UEP requirements by virtue of its inherent hierarchy, this per-subcarrier pairing forms a natural, engineering-effective modeling basis. It inherits the original UEP design logic and generalizes it from single-carrier MIMO subchannels to multi-carrier MIMO-OFDM subcarriers for practical multi-path fading scenarios.

As shown in Fig. \ref{fig_1}(b), the video context $\mathbf{c}_t$ and sampled subcarrier channels $\mathbf{h}_t$ are mapped to the identical feature space through respective encoders and normalization with the cosine similarity computation afterwards. The formulation of the context-subcarrier correlation map is similar to \cite{cvst} as
\begin{align}
	m_{t,ij}(\mathbf{c}_{t},\mathbf{h}_{t}) = \frac{\exp(sim(\mathbf{V}_{\theta_1}(\mathbf{c}_{t,i}),\mathbf{V}_{\theta_2}(\mathbf{h}_{t,j}))/\tau)}{\sum_{j=1}^{N_s/m_h}\exp(sim(\mathbf{V}_{\theta_1}(\mathbf{c}_{t,i}),\mathbf{V}_{\theta_2}(\mathbf{h}_{t,j}))/\tau)},
\end{align}
where $m_{t,ij}(\mathbf{c}_{t},\mathbf{h}_{t})$ represents the score for providing the relative ranking of matched context-subcarrier pair with the $i$-th context group, the $j$-th subcarrier group, $\mathbf{V}_{\theta_1}(\cdot)$ and $\mathbf{V}_{\theta_2}(\cdot)$ encapsulate the corresponding feature encoder and normalization process. $sim(\cdot)$ represents the cosine similarity computation, $\tau$ is the learnable temperature parameter.
\vspace{-5pt}
\section{Time-Correlated Variable Length Coding}

\begin{figure*}[htbp]
	\centering
	\includegraphics[width=6.3in]{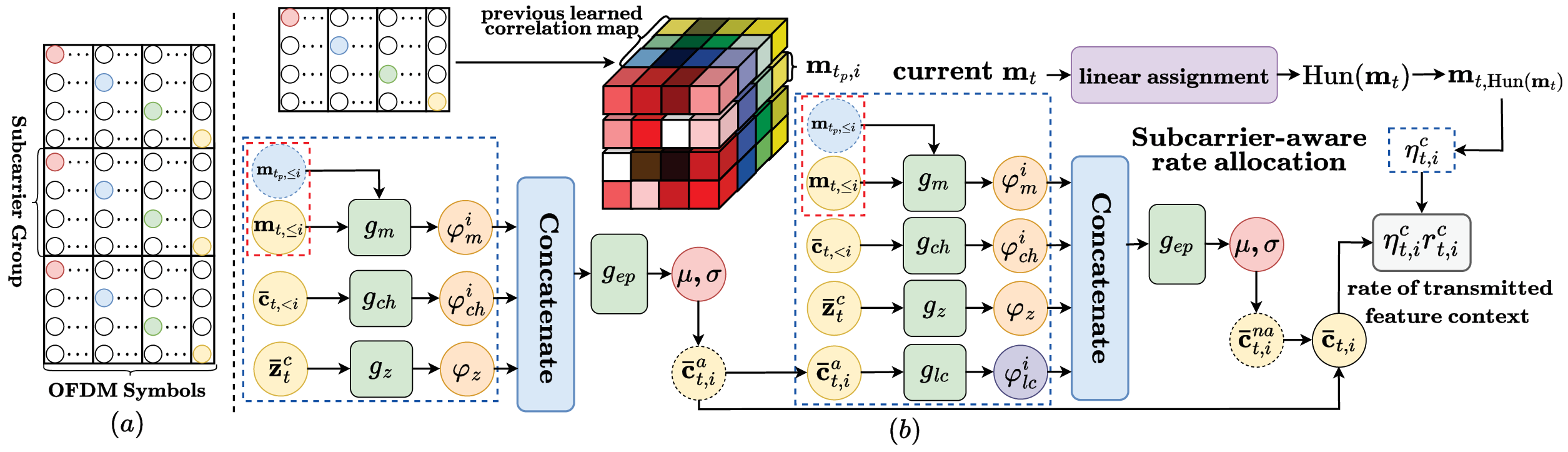}
	\vspace{-5pt}
	\caption{(a) The recursive subcarrier sampling for time-correlated channels across OFDM symbols. (b) The structure of time-correlated multi-reference variable length coding.}
	\vspace{-15pt}
	\label{fig_2}
\end{figure*}

To further exploit the time correlation of multi-path channels, we thus propose the periodic subcarrier sampling method and its corresponding reference embeddings.
\vspace{-10pt}
\subsection{Recursive Subcarrier Sampling}

As illustrated in Sec. II, periodic subcarrier sampling is employed to reduce the computational overhead of CSI acquisition and precoding, wherein a single sampled subcarrier represents a block of adjacent ones. However, the inherent time correlation of wireless channels motivates the use of historically sampled data within each subcarrier group to better infer current channel conditions. In this way, we thus propose a recursive subcarrier sampling method for time-correlated channels. As shown in Fig. \ref{fig_2}(a), within the same OFDM symbol, the same relative position is sampled across all subcarrier groups. In the subsequent symbol, the sampling position is shifted uniformly by one index in every group. Such a cyclic pattern ensures sampling all subcarriers over $m_h$ consecutive symbols, which means that previously adjacent subcarriers can be exploited for current channel condition understanding during video frame transmission.
\vspace{-10pt}
\subsection{Time-Correlated Reference Embedding for Entropy Coding}

Following the checkerboard-based NTC entropy coding \cite{cvst}, multiple references are embedded as supplemental enhancement information (SEI) into the hyperprior entropy model for rate adjustment and unequal error protection. To exploit the adjacent subcarrier information, we collected the previously learned correlation maps $\mathbf{m}_s$ based on the corresponding sampled subcarrier groups as time-correlated references. The modified entropy model is formulated as
\begin{equation}
	\begin{aligned}
		&P_{\tilde{\mathbf{c}}_{t}^{a}|\mathbf{m}_s,\tilde{\mathbf{z}}_t^c}(\tilde{\mathbf{c}}_{t}^{a}|\mathbf{m}_s,\tilde{\mathbf{z}}_t^c)\\&=\prod_{i}P_{\tilde{\mathbf{c}}_{t,i}^{a}|\mathbf{m}_{s,i},\tilde{\mathbf{z}}_t^c,\tilde{\mathbf{c}}_{t,<i}}(\tilde{\mathbf{c}}_{t,i}^{a}|\mathbf{m}_{s,i},\tilde{\mathbf{z}}_t^c,\tilde{\mathbf{c}}_{t,<i})\\&=\prod_{i}(\mathcal{L}(\tilde{\mu}_{t,i}^{c,a},\tilde{\sigma}_{t,i}^{c,a})*\mathcal{U}(-\frac{1}{2},\frac{1}{2}))(\tilde{\mathbf{c}}_{t,i}^a),
	\end{aligned}
\end{equation}
\begin{equation}
	\begin{aligned}
		&P_{\tilde{\mathbf{c}}_{t}^{na}|\mathbf{m}_s,\tilde{\mathbf{z}}_t^c,\tilde{\mathbf{c}}_{t}^a}(\tilde{\mathbf{c}}_{t}^{na}|\mathbf{m}_s,\tilde{\mathbf{z}}_t^c,\tilde{\mathbf{c}}_{t}^a)\\&=\prod_{i}P_{\tilde{\mathbf{c}}_{t,i}^{na}|\mathbf{m}_{s,i},\tilde{\mathbf{z}}_t^c,\tilde{\mathbf{c}}_{t,<i},\tilde{\mathbf{c}}_{t,i}^a}(\tilde{\mathbf{c}}_{t,i}^{na}|\mathbf{m}_{s,i},\tilde{\mathbf{z}}_t^c,\tilde{\mathbf{c}}_{t,<i},\tilde{\mathbf{c}}_{t,i}^a)\\&=\prod_{i}(\mathcal{L}(\tilde{\mu}_{t,i}^{c,na},\tilde{\sigma}_{t,i}^{c,na})*\mathcal{U}(-\frac{1}{2},\frac{1}{2}))(\tilde{\mathbf{c}}_{t,i}^{na}),
	\end{aligned}
\end{equation}
where $\mathbf{m}_s=\mathrm{Con}(\mathbf{m}_{t_p},\mathbf{m}_t)$, $t_p=[t-(t \mod \frac{N_s}{m_h}),\cdots,t-1]$ represents previously learned context-subcarrier correlation maps, ${(\cdot\mod\cdot)}$ is the modulo operation, $\mathrm{Con}(\cdot,\cdot)$ is the concatenation function. $\tilde{\mathbf{z}}_t^c$ is the uniformly-noised hyperprior parameter, $\tilde{\mathbf{c}}_{t,i}$ is the quantized representation, which is modeled as Laplace distribution. $\tilde{\mathbf{c}}_t^a$ and $\tilde{\mathbf{c}}_t^{na}$ divide $\tilde{\mathbf{c}}_t$ into the anchored part and non-anchored part, respectively.

The learned mean and variance are given as
\begin{equation}
	\begin{aligned}
		(\tilde{\mu}_{t,i}^{c,a},\tilde{\sigma}_{t,i}^{c,a})&=g_{ep}(\mathbf{\varphi}_m^i,\mathbf{\varphi}_{ch}^i,\mathbf{\varphi}_z)\\&=g_{ep}(g_m(\mathbf{m}_{s,\le i}),g_{ch}(\mathbf{\tilde{c}}_{t,\le i}),g_z(\mathbf{\tilde{z}}_t)),
	\end{aligned}
\end{equation}
\begin{equation}
	\begin{aligned}
		(\tilde{\mu}_{t,i}^{c,na},\tilde{\sigma}_{t,i}^{c,na})&=g_{ep}(\mathbf{\varphi}_m^i,\mathbf{\varphi}_{ch}^i,\mathbf{\varphi}_z,\mathbf{\varphi}_{lc}^i)\\&=g_{ep}(g_m(\mathbf{m}_{s,\le i}),g_{ch}(\mathbf{\tilde{c}}_{t,\le i}),g_z(\mathbf{\tilde{z}}_t),g_{lc}(\mathbf{\tilde{c}}_{t,i}^a)),
	\end{aligned}
\end{equation}
where $g_m(\cdot)$, $g_{ch}(\cdot)$, $g_z(\cdot)$, and $g_{lc}(\cdot)$ are the corresponding reference generators, $\mathbf{\varphi}_m^i$, $\mathbf{\varphi}_{ch}^i$, $\mathbf{\varphi}_z$, and $\mathbf{\varphi}_{lc}^i$ are the references for entropy coding.

After that, non-parametric fully factorized density is utilized to model the hyperprior distribution as
\begin{align}
	P_{\tilde{\mathbf{z}}_{t}^c}(\tilde{\mathbf{z}}_{t}^c)=\prod_{j}(P_{\mathbf{z}_{t,j}^c|\psi^{(j)}}(\mathbf{z}_{t,j}^c|\psi^{(j)})*\mathcal{U}(-\frac{1}{2},\frac{1}{2}))(\tilde{\mathbf{z}}_{t,j}^c),
\end{align}
where $\psi^{(j)}$ encapsulates all the parameters of $P_{\mathbf{z}_{t,j}^c|\psi^{(j)}}$.

With the learned entropy model, the allocated channel bandwidth cost for the feature context is formulated as
\begin{equation}
	\begin{aligned}
		k_{t,i}^c&=\eta_{t,i}^cr_{t,i}^c\\&=-\eta_{t,i}^c(\log P_{\bar{\mathbf{c}}_{t,i}^{a}|\mathbf{m}_{s,\le i},\bar{\mathbf{z}}_t^c,\bar{\mathbf{c}}_{t,<i}}(\bar{\mathbf{c}}_{t,i}^{a}|\mathbf{m}_{s,\le i},\bar{\mathbf{z}}_t^c,\bar{\mathbf{c}}_{t,<i})\\&+\log P_{\bar{\mathbf{c}}_{t,i}^{na}|\mathbf{m}_{s,\le i},\bar{\mathbf{z}}_t^c,\bar{\mathbf{c}}_{t,<i},\bar{\mathbf{c}}_{t,i}^a}(\bar{\mathbf{c}}_{t,i}^{na}|\mathbf{m}_{s,\le i},\bar{\mathbf{z}}_t^c,\bar{\mathbf{c}}_{t,<i},\bar{\mathbf{c}}_{t,i}^a)),
	\end{aligned}
\end{equation}
where $\eta_{t,i}^c$ is the $i$-th group rate adjustment hyperparameter.

Then, the total channel bandwidth cost is collected as
\begin{equation}
	\begin{aligned}
		k_{t}^c&=-\sum_{i}\eta_{t,i}^c(\log P_{\bar{\mathbf{c}}_{t,i}^{a}|\mathbf{m}_{s,\le i},\bar{\mathbf{z}}_t^c,\bar{\mathbf{c}}_{t,<i}}(\bar{\mathbf{c}}_{t,i}^{a}|\mathbf{m}_{s,\le i},\bar{\mathbf{z}}_t^c,\bar{\mathbf{c}}_{t,<i})\\&+\log P_{\bar{\mathbf{c}}_{t,i}^{na}|\mathbf{m}_{s,\le i},\bar{\mathbf{z}}_t^c,\bar{\mathbf{c}}_{t,<i},\bar{\mathbf{c}}_{t,i}^a}(\bar{\mathbf{c}}_{t,i}^{na}|\mathbf{m}_{s,\le i},\bar{\mathbf{z}}_t^c,\bar{\mathbf{c}}_{t,<i},\bar{\mathbf{c}}_{t,i}^a)).
	\end{aligned}
\end{equation}

Finally, the transmission cost for M-CVST is formulated as
\begin{align}
	k_{t}=k_{t}^c + k_{t}^v + k_{t}^{c_z} + k_{t}^{v_z},
\end{align}
where $k_{t}^v$ and $k_{t}^c$ are the bandwidth cost of $\mathbf{v}_t$ and $\mathbf{c}_t$, $k_{t}^{c_z}$ and $k_{t}^{v_z}$ are the hyperprior vector transmission bandwidth cost.

The illustration of proposed time-correlated multi-reference variable length coding is shown in Fig. \ref{fig_2}(b). $\mathbf{\varphi}_m^i$ encapsulates both the previous and current channel information. In this way, the training loss is defined as
\begin{align}
	L_{t}&=k_t+\lambda\cdot(D_t(\mathbf{x}_t,\mathbf{\hat{x}}_t)+D_t(\mathbf{x}_t,\mathbf{\bar{x}}_t)),
\end{align}
where $\lambda$ is the Lagrange multiplication from a set of predefined $\lambda$ values for variable rate coding. $D_t(\mathbf{x}_t,\mathbf{\hat{x}}_t)$ is the frame reconstruction loss. $D_t(\mathbf{x}_t,\mathbf{\bar{x}}_t)$ is the NTC loss which performs as a reweighting term for keeping training stable.

\vspace{-5pt}
\section{Numerical Results}
In this section, numerical results are presented to verify the effectiveness of M-CVST. 
\vspace{-10pt}
\begin{figure*}[htbp]
	\centering  %图片全局居中
	\hspace{-20mm}
	\subfigure[PNSR results.]{
		\includegraphics[width=0.29\linewidth]{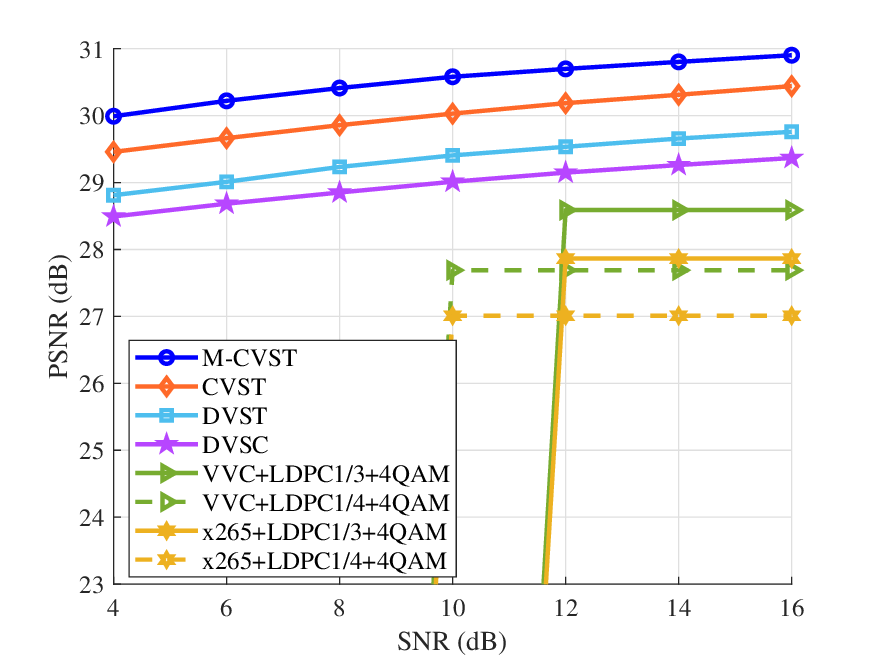}}
	\hspace{-8mm}
	\subfigure[MS-SSIM results.]{
		\includegraphics[width=0.29\linewidth]{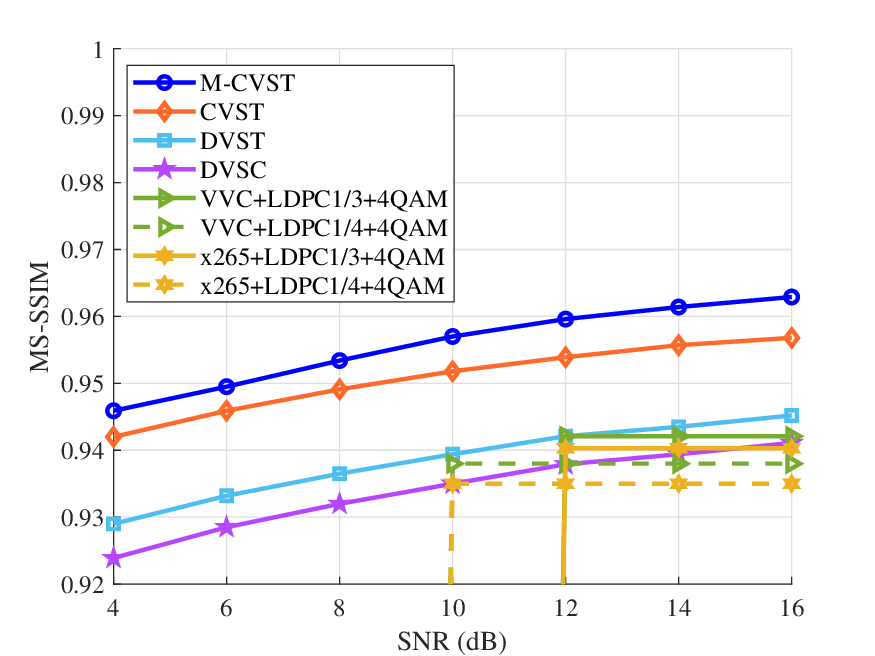}}
	\hspace{-8mm}
	\subfigure[LPIPS results.]{
		\includegraphics[width=0.29\linewidth]{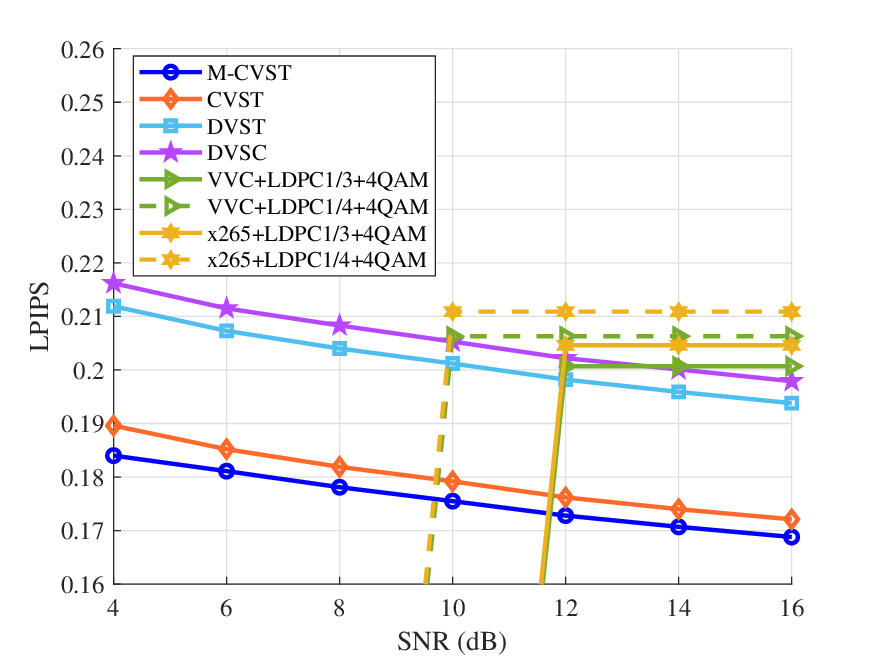}}
	\hspace{-8mm}
	\subfigure[PNSR results.]{
		\includegraphics[width=0.29\linewidth]{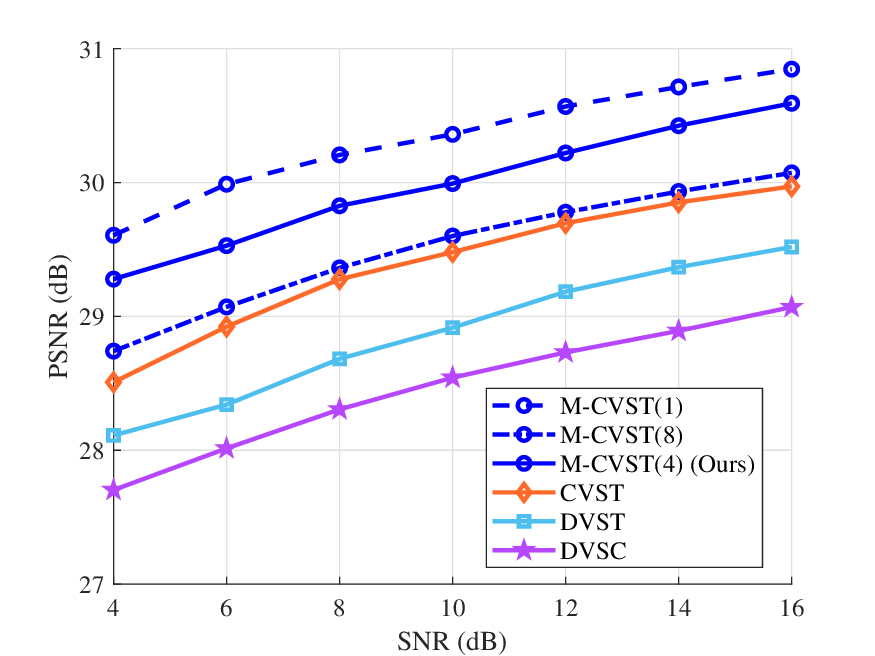}}
	\hspace{-20mm}
	\vspace{-5pt}
	\caption{(a)-(c) Quality of the reconstructed images versus the SNRs under MIMO CDL channels (G1) (d) (G2) ($R$ = 0.0347).}
	\vspace{-15pt}
	\label{fig_3}
\end{figure*}
\subsection{Experimental Setups}

\subsubsection{Datasets}

We quantify the performances of M-CVST over Vimeo-90k dataset and HEVC dataset \cite{hevc}. During model training, video frames are randomly cropped to 256$\times$256$\times$3. While for model testing, we adapt HEVC ClassC dataset (832 × 480) \cite{hevc}. The channel configuration is set in Tab. \ref{table1}. G1 and G2 are the defined channel configuration groups. 'symbol/frame' is the OFDM symbol number within a single frame transmission process. Similar to the division in \cite{LCFSC}, MIMO CSI matrices are generated according to \cite{sionna} with 1000 OFDM symbols for training and 100 symbols for testing. 

\begin{table}[htbp]
	\centering
	\caption{MIMO CDL channel configuration (G1, G2)}
	\label{table1}
	\small
	\begin{tabular}{|c|c|c|c|c|}  
		\hline
		& & & &\\[-6pt]  
		&MIMO&8$\times$8&Subcarrier&64 \\
		\hline
		& & & &\\[-6pt]  
		&\thead{Channel \\ Model}&\thead{3GPP 38.901 \\ CDL-C \cite{sionna}}&Carrier Frequency&2.6 GHz \\
		\hline
		& & & &\\[-6pt] 
		G1&Speed&40 km/h&symbol/frame&1\\
		\hline
		& & & &\\[-6pt] 
		G2&Speed&80 km/h&symbol/frame&4\\
		\hline
	\end{tabular}
\end{table}

\subsubsection{Model Deployment Details}
The network deployment of M-CVST is the same as \cite{cvst}. Feature channel dimension $L$ is set as 64. While $m_h=8$ and $m_c=8$. We assume that one OFDM symbol period covers a single frame transmission. The SNR set is defined as [0, 2, 4, 6, 8, 10, 12, 14] dB while $\lambda$ set is defined as [0.015, 0.06, 0.12, 0.20, 0.32]. During training, SNR and $\lambda$ are randomly selected for adapting variable CBRs and SNRs.

\subsubsection{Comparison Benchmarks}
In the experiments, several benchmarks are given as below

$\textbf{M-CVST ($m$)}$: M-CVST samples one subcarrier for every consecutive $m$ subcarriers.

$\textbf{CVST}$: The context-aware wireless video transmission framework \cite{cvst} with variable length and rate coding which can be assumed as the ablation benchmark of M-CVST without time-correlated reference embeddings.

$\textbf{DVSC}$: The DL-empowered deep video transmission framework \cite{dvsc} with SNR-adaptive channel coder and semantic restoration at the receiving end.

$\textbf{DVST}$: The wireless video semantic transmission framework \cite{dvst} with rate-adaptive contextual transmission.

$\textbf{VVC/x265+LDPC+QAM}$: The SSCC scheme with VVC \cite{vvenc}/x265 \cite{ffmpeg} video codec and 5G LDPC \cite{sionna}, along with the quadrature amplitude modulation (QAM). SVD precoding, random interleave (RI) method and waterfilling (WF) power allocation are also adapted.

\subsubsection{Evaluation Metrics}

We leverage the pixel-wise metric peak signal-to-noise ratio (PSNR) and the perceptual-level multi-scale structural similarity (MS-SSIM) along with learned perceptual image patch similarity (LPIPS) as measurements for the reconstructed image quality. According to \cite{dvst}, CBR is employed to evaluate compression performance as
\begin{align}
	CBR=\frac{\sum_{t=1}^{T}k_{t}}{T\times H\times W\times3}.
\end{align}

\begin{figure*}[htbp]
	\centering  %图片全局居中
	\subfigure[PNSR for the reconstructed images.]{
		\includegraphics[width=0.30\linewidth]{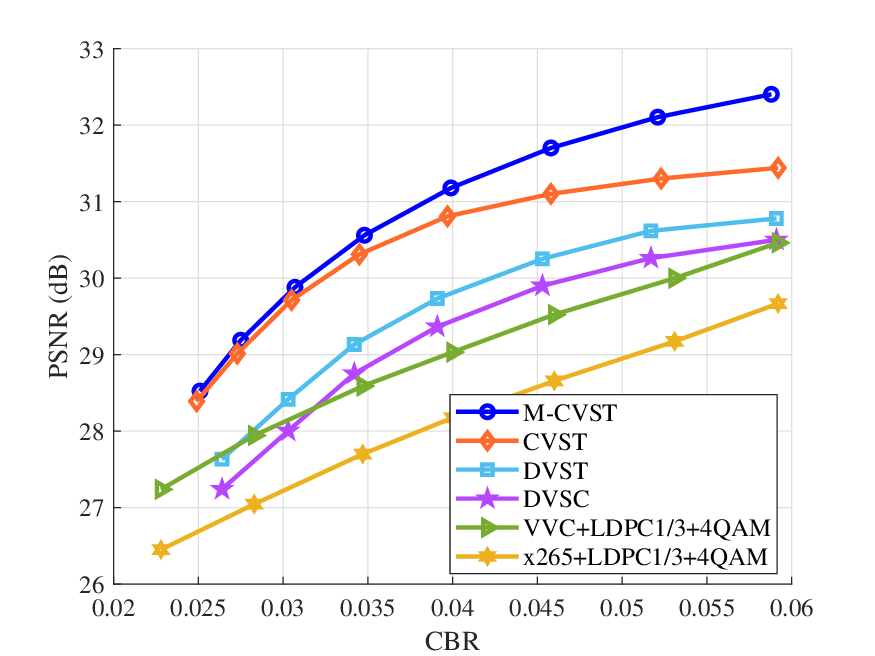}}
	\subfigure[MS-SSIM for the reconstructed images.]{
		\includegraphics[width=0.30\linewidth]{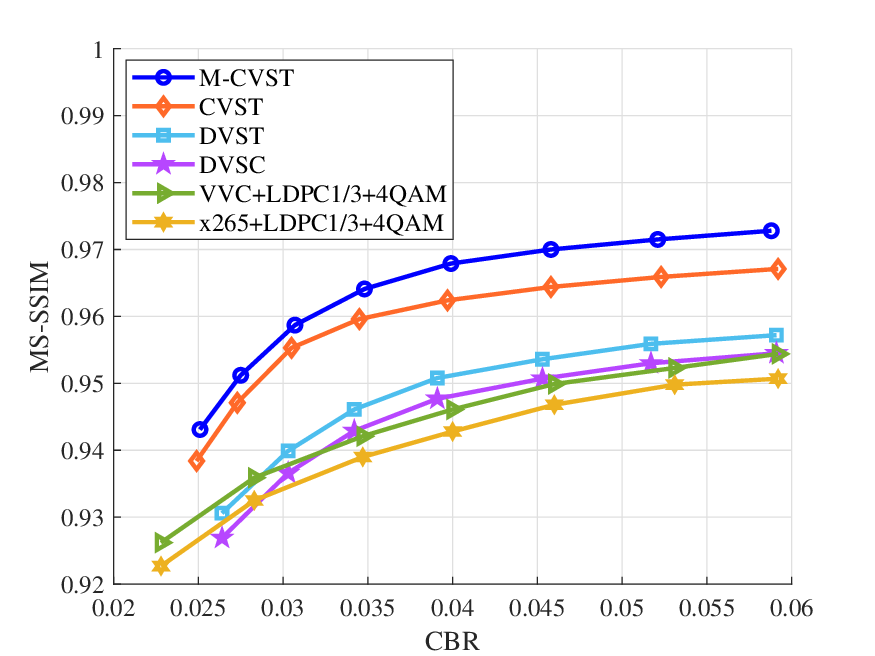}}
	\subfigure[LPIPS for the reconstructed images.]{
		\includegraphics[width=0.30\linewidth]{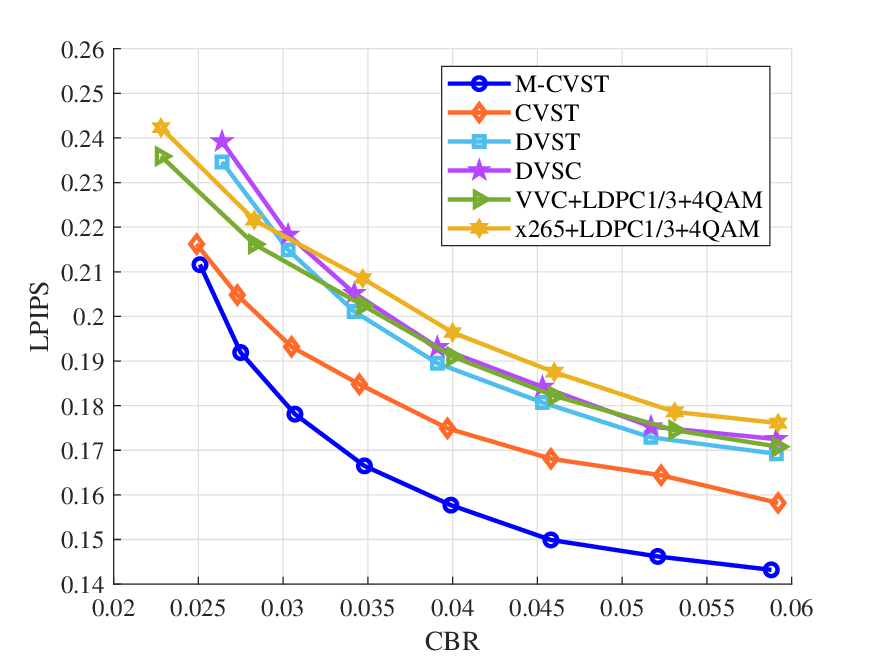}}
	\vspace{-5pt}
	\caption{Quality of the reconstructed images versus the CBRs under MIMO CDL channels (G1, SNR = 12 dB).}
	\vspace{-15pt}
	\label{fig_4}
\end{figure*}
\vspace{-20pt}
\subsection{Results Analysis}

\subsubsection{SNR Performances}
We first evaluate the anti-noise performance of M-CVST over MIMO CDL channels under fixed CBRs, using a single model with perfect CSI tested across multiple SNRs. As shown in Fig. \ref{fig_3}, M-CVST outperforms all benchmarks significantly. It outperforms DVST and DVSC by approximately 1.5 dB in PSNR, validating the robustness of context-aware coding with multi-reference variable-length coding. Compared with CVST, M-CVST achieves further gains owing to its time-correlation-aware designs in SEI embedding and multi-reference entropy coding. Against traditional VVC+LDPC+QAM schemes, M-CVST delivers much higher performance and stability, avoiding the cliff effect in harsh channel conditions. Perceptual metrics including MS-SSIM and LPIPS in Fig. \ref{fig_3}(b)–(c) further confirm the superiority of M-CVST over multi-path MIMO channels.

While for Fig. \ref{fig_3}(d), some ablation study is conducted. MCVST (1) refers to sampling full CSI and feeding it back to the transmitter for precoding, which is assumed as the common upper bound. For G2 channel condition, we employ interval 4 for recursive subcarrier sampling. It is observed that the performance gap between MCVST (1) and MCVST (4) is even larger than the gap between MCVST (4) and MCVST (8), which demonstrates the efficiency of proposed recursive sampling in terms of performance-complexity tradeoff.
\vspace{-2pt}
\subsubsection{CBR Performances}
We then evaluate M-CVST's bandwidth compression performance at SNR = 12 dB. As shown in Fig. \ref{fig_4}, M-CVST consistently achieves significant performance gains over all comparative schemes. Notably, the performance gaps between M-CVST and other DL-based schemes widen as CBR increases. This advantage stems primarily from the proposed time-correlated multi-reference variable length coding module whose well-designed CSI and CBR-aware SEI embeddings enable efficient variable length coding across diverse rate points within a single model. Furthermore, M-CVST outperforms traditional VVC-based schemes (VVC+LDPC+QAM), demonstrating the superior compression efficiency of its jointly optimized context transmission and entropy coding. To conclude, M-CVST maintains stable performance gains across varying video content types and motion complexities, demonstrating its flexibility and robustness in diverse video transmission scenarios.

\begin{table}[htbp]
	\centering
	\caption{Evaluation of complexity and computation cost.}
	\label{table2}
	\small
	\begin{tabular}{|c|c|c|}  
		\hline 
		& &\\[-6pt] 
		Metric&FLOPs (G)&Throughput (image/sec) \\
		\hline
		& &\\[-6pt]  
		M-CVST (full CSI)&362.67&6.71 \\
		\hline
		& &\\[-6pt]  
		M-CVST (Ours)&384.10&8.37 \\
		\hline
		& &\\[-6pt]  
		CVST&387.41&8.76 \\
		\hline
		& &\\[-6pt] 
		VVC+LDPC+QAM &/&4.31 \\
		\hline
	\end{tabular}
\end{table}
\subsubsection{Computation Complexity}
Finally, we analyze M-CVST's computational cost. In Tab. \ref{table2}, M-CVST achieves competitive throughput while delivering substantially better reconstruction quality. Although the proposed multi-reference entropy coding introduces additional parameters, M-CVST retains a practical inference speed (comparable to CVST). In contrast, M-CVST with full feedback CSI introduces much computation cost and CSI feedback overhead; Traditional VVC+LDPC+QAM incurs low throughput due to its time-consuming rate-distortion optimization process for every block. Overall, these results verify that the proposed M-CVST along with recursive subcarrier sampling achieves a superior performance–complexity tradeoff with affordable computation cost and improved efficiency, making it suitable for practical wireless transmission systems.

\vspace{-12pt}

\section{Conclusion}
This paper presents M-CVST for multi-path MIMO-OFDM channels, featuring a subcarrier-level context-correlation map, recursive subcarrier sampling and time-correlated embedding. Extensive experiments verify superior performance of M-CVST over existing semantic and traditional SSCC schemes under varying SNRs and CBRs, with robust anti-noise and compression efficiency. Future work will explore more channel scenarios and multi-user deployments for practical application.

%\vspace{-10pt}
\fontsize{8pt}{10pt}\selectfont
%\bibliographystyle{IEEEtran}
%\bibliography{Ref}

\begin{thebibliography}{50}
%\bibliographystyle{IEEEtran}
\vspace{-2pt}
\bibitem{265}
G. Sullivan, J. Ohm, W. Han, and T. Wiegand, ``Overview of the high efficiency video coding (HEVC) standard," \emph{IEEE Trans. Circuits Syst. Video Technol.}, vol. 22, no. 12, pp. 1649-1668, Dec. 2012.

\bibitem{vvc}
B. Benjamin, et al., ``Overview of the versatile video coding (VVC) standard and its applications," \emph{IEEE Trans. Circuits Syst. Video Technol.}, vol. 31, no.10, pp. 3736-3764, Aug. 2021.

\bibitem{wvscd}
B. Xie et al., ``Wireless Video Semantic Communication with Decoupled Diffusion Multi-frame Compensation," \emph{IEEE Trans. Commun.}, vol. 74, pp. 987-1002, Nov. 2025.

\bibitem{cra}
Z. Zhao et al., ``Compression Ratio Allocation for Probabilistic Semantic Communication With RSMA," \emph{IEEE Trans. Commun.}, vol. 73, no. 9, pp. 7304-7318, Sept. 2025.

\bibitem{dvsc}
H. Niu, L. Wang, Z. Lu, K. Du, and X. Wen, ``Deep learning enabled video semantic transmission against multi-dimensional noise," in \emph{Proc. IEEE Glob. Commun. Conf.  Workshops (GLOBECOM Workshops)}, Kuala Lumpur, Malaysia, pp. 1267-1272, Dec. 2023.

\bibitem{dvst}
S. Wang et al., ``Wireless Deep Video Semantic Transmission," \emph{IEEE J. Select. Areas Commun.}, vol. 41, no. 1, pp. 214-229, Jan. 2023.

\bibitem{cvst}
B. Xie et al., ``Context Video Semantic Transmission with Variable Length and Rate Coding over MIMO Channels," Dec. 2025. [Online]. Available: \url{https://arxiv.org/abs/2601.06059}.

\bibitem{hevc}
F. Bossen et al., ``Common Test Conditions and Software Reference Configurations," document JCTVC-L1100, vol. 12, no. 7, 2013.

\bibitem{LCFSC}
B.~Xie, Y.~Wu, Y.~Shi, W.~Zhang, S.~Cui, and M.~Debbah, ``Robust image semantic coding with learnable CSI fusion masking over MIMO fading channels," \emph{IEEE Trans. Wireless Commun.}, vol. 23, no. 10, pp. 14155-14170, Oct. 2024.

\bibitem{sionna} 
H., Jakob, et al., ``Sionna: An open-source library for next-generation physical layer research," Mar. 2022. [Online]. Available: \url{https://arxiv.org/abs/2203.11854}.

\bibitem{vvenc}
W. Adam, et al., ``VVenC: An open and optimized VVC encoder implementation," in \emph{IEEE Int. Conf. Multimedia Expo Workshops}, Shenzhen, China, Jun. 2021.

\bibitem{ffmpeg}
S. Tomar, ``Converting video formats with FFmpeg," \emph{Linux J.}, vol. 2006, no. 146, Jun. 2006.

\end{thebibliography}

\end{document}